\documentclass[twocolumn,notitlepage,footinbib,amsmath,amssymb,amsfonts]{revtex4-1}

\usepackage{graphicx}
\usepackage{multirow}
\usepackage{color}
\usepackage{enumerate}
\usepackage[normalem]{ulem}
\usepackage{ytableau}
\usepackage[hypertexnames=false]{hyperref}

\newcommand{\mb}[1]{\ensuremath{\mathbf{#1}}}

\newcommand{\mr}[1]{\ensuremath{\mathrm{#1}}}
\newcommand{\mbb}[1]{\ensuremath{\mathbb{#1}}}

\newcommand{\im}{\ensuremath{\mathrm{Im}}}
\newcommand{\la}{\ensuremath{\langle}}
\newcommand{\ra}{\ensuremath{\rangle}}

\newcommand{\ket}[1]{\ensuremath{| #1 \rangle}}

\newcommand{\matel}[3]{\ensuremath{\langle #1 | #2 | #3 \rangle}}

\newcommand{\Eq}[1]{Eq.~\eqref{#1}}

\newcommand{\Fig}[1]{Fig.~\ref{#1}}

\newcommand{\Sec}[1]{Sec.~\ref{#1}}

\newcommand{\Ref}[1]{Ref.~\cite{#1}}

\allowdisplaybreaks

\begin{document}

\title{Doublon-holon origin of the subpeaks at the Hubbard band edges}
\author{Seung-Sup B. Lee}
\author{Jan von Delft}
\author{Andreas Weichselbaum}
\affiliation{Physics Department, Arnold Sommerfeld Center for Theoretical Physics and Center for NanoScience, Ludwig-Maximilians-Universit\"{a}t M\"{u}nchen, Theresienstra{\ss}e 37, 80333 M\"{u}nchen, Germany}
\date{\today}
\begin{abstract}
Dynamical mean-field theory (DMFT) studies frequently observe a fine structure in the local spectral function of the SU(2) Fermi-Hubbard model at half filling:
in the metallic phase close to the Mott transition,
subpeaks emerge at the inner edges of the Hubbard bands.
Here we demonstrate that these subpeaks originate from the low-energy effective interaction of doublon-holon pairs,
by investigating how the correlation functions of doublon and holon operators contribute to the subpeaks.
A mean-field analysis of the low-energy effective Hamiltonian provides results consistent with our DMFT calculation using the numerical renormalization group as an impurity solver. In the SU(3) and SU(4) Hubbard models, the subpeaks become more pronounced due to the increased degeneracy of doublon-holon pair excitations.
\end{abstract}

\maketitle

{\it Introduction.---}
Dynamical mean-field theory (DMFT) \cite{Georges1996,*Kotliar2006} provides a widely successful
approach in understanding strongly correlated systems.
It treats a lattice problem by self-consistently solving
an effective impurity model whose impurity and bath correspond
to a lattice site and the rest of the lattice, respectively.
Thus the performance of DMFT calculations directly depends
on which particular impurity solver is chosen.

A benchmark calculation for various impurity solvers
is the paramagnetic Mott transition in the half-filled
$\mr{SU}(2)$ Hubbard model at temperature $T = 0$
which is characterized by a striking change in the
local spectral functions \cite{Zhang1993,Bulla1999}.
In the metallic phase, the spectral function features a
quasiparticle peak (QP) at the Fermi level, and two Hubbard bands (HBs)
below and above the Fermi level each.
In the insulating phase, the QP disappears
and a gap opens between two HBs.

In the metallic phase close to the transition,
many DMFT studies have observed sharp subpeaks
that emerge at the inner edges of the HBs,
by using different real-frequency impurity solvers:
perturbative methods~\cite{Zhang1993},
the density-matrix renormalization group
(DMRG)~\cite{Karski2005,Karski2008,Ganahl2014,*Ganahl2015,Wolf2014},
the numerical renormalization group (NRG)~\cite{Zitko2009},
and exact diagonalization~\cite{Granath2012,Lu2014}.
In contrast, quantum Monte Carlo solvers,
which obtain the spectral functions on the real frequency axis
via (numerically ill-posed) analytic continuation,
have not found these subpeaks.
The subpeaks give rise to
distinct features in the momentum-resolved spectral function~\cite{Karski2008},
measurable by photoemission spectroscopy~\cite{Mo2003,Sekiyama2004}.
Despite these frequent consistent observations,
the physical origin of the subpeaks and their relevance in more general
(e.g., multi-flavor) models remained unclear.

In this Letter, we show that the subpeaks are induced
by the effective doublon-holon (DH)~\cite{Yokoyama2006,*Phillips2010,*Sato2014,*Zhou2014,*Prelovsek2015,*Zhou2016} pair interaction
originating from a second-order virtual process,
where a doublon (holon) means an excitation
that one particle is added to (removed from) a lattice site
with average integer filling.
We compute the correlation functions of doublon and holon operators in the $\mr{SU}(2)$ Hubbard model,
by using DMFT with NRG~\cite{Wilson1975,Bulla2008} as an impurity solver,
and demonstrate that these correlation functions manifest the peak structure associated with the subpeaks.
We reproduce the peak structure of doublon and holon correlators via a mean-field analysis of the low-energy effective Hamiltonian
obtained by a generalized Schrieffer-Wolff transformation (SWT)~\cite{Bukov2016,*Bukov2015,Lee2017a}.
Both approaches consistently result in a linear dependence of the subpeak position vs.\ interaction strength.
From our DMFT+NRG calculations of general $\mr{SU}(N)$ Hubbard models
for $N = 2, 3, 4$,
we observe that the subpeaks become more pronounced with increasing $N$,
since the DH pair excitations become more degenerate due to
the larger $\mr{SU}(N)$ symmetry.

{\it System.---}
The $\mr{SU}(N)$ Hubbard model describes $N$ flavors of fermions
on a lattice with local repulsive interactions,
recently realized in ultracold atom experiments with tunable
$N$ \cite{Taie2012,*Hofrichter2016}.
The hopping amplitude $v$, the interaction strength $U$, and the chemical potential are flavor-independent, thus the system has $\mr{SU}(N)$ flavor symmetry.
Its Hamiltonian is $H = H_U + H_v + H_\mu$, where
$H_U = \frac{U}{2} \sum_i (\hat{n}_i - \bar{n})^2$,
$H_v = v \sum_{\la i,j \ra, \nu} c_{i\nu}^\dagger c_{j\nu} + \text{h.c.}$,
and $H_\mu = - \mu \sum_i \hat{n}_i$.
Here $c_{i\nu}$ annihilates a particle of flavor $\nu = 1, \ldots, N$ at lattice site $i$,
$\hat{n}_i = \sum_{\nu} c_{i\nu}^\dagger c_{i\nu}$ is the particle number operator at site $i$,
$\la i,j \ra$ indicates nearest neighbours,
$\bar{n}$ is a parameter for the desired average occupation,
and $\mu$ is a fine tuning of chemical potential to
achieve $\la \hat{n}_i \ra = \bar{n}$.
Throughout this paper, we focus on $T = 0$ and the average occupation number as an integer closest to half filling $\bar{n} = \lfloor N/2 \rfloor$,
by fixing $\mu = 0$ for $N = 2, 4$, and 
fine-tuning $\mu$ for $N = 3$.

{\it Doublon and holon.---}
For integer average occupation $\bar{n}$,
we define doublon and holon creation operators as
\begin{equation}
d_{i\nu}^\dagger \equiv P_{i, \bar{n}+1} c_{i\nu}^\dagger\,, \,\,\, h_{i\nu}^\dagger \equiv P_{i, \bar{n}-1} c_{i\nu} \,,
\label{eq:DHdef}
\end{equation}
where $P_{in}$ means the projector onto the subspace
in which the site $i$ has $n$ particles.
For the $\mr{SU}(2)$ case, at half filling, $\bar{n} = 1$,
these operators reduce to $d_{i\nu} = c_{i\nu} \hat{n}_{i\bar{\nu}}$, $h_{i\nu} = c_{i\nu}^\dagger (1 - \hat{n}_{i\bar{\nu}})$ with $\hat{n}_{i\nu} = c_{i\nu}^\dagger c_{i\nu}$ and $\bar{\nu} = 3 - \nu$,
and they completely constitute the particle operator $c_{i\nu} = d_{i\nu} + h_{i\nu}^\dagger$.
Then the particle correlator can be decomposed into four
doublon and holon correlators,
$A_{c c^\dagger} (\omega) = A_{d d^\dagger} + A_{d h} + A_{h^\dagger d^\dagger} + A_{h^\dagger h}$,
where $A_{X Y}(\omega) \equiv \tfrac{-1}{\pi} \im G_{X Y}$,
with $G_{X Y}(t) = -i\vartheta(t) \la [ X (t), Y (0) ]_\pm \ra_T$ being
the retarded correlation function 
of the fermionic ($+$) or bosonic ($-$) local operators $X$ and $Y$
acting on the same site.
In the particle-hole symmetric case, only two correlators
are independent:
``diagonal'' correlators
$A_{d d^\dagger} (\omega) = A_{h^\dagger h} (-\omega)$
which are asymmetric,
and ``off-diagonal'' correlators
$A_{h^\dagger d^\dagger} (\omega) = A_{d h} (\omega)$
which are symmetric under $\omega \leftrightarrow -\omega$.
For $N > 2$ flavors,
the decomposition of $c_{i\nu}$ acquires
more terms than $d_{i\nu}$ and $h_{i\nu}^\dagger$~\cite{Lee2017a}.

\begin{figure}
\centerline{\includegraphics[width=.47\textwidth]{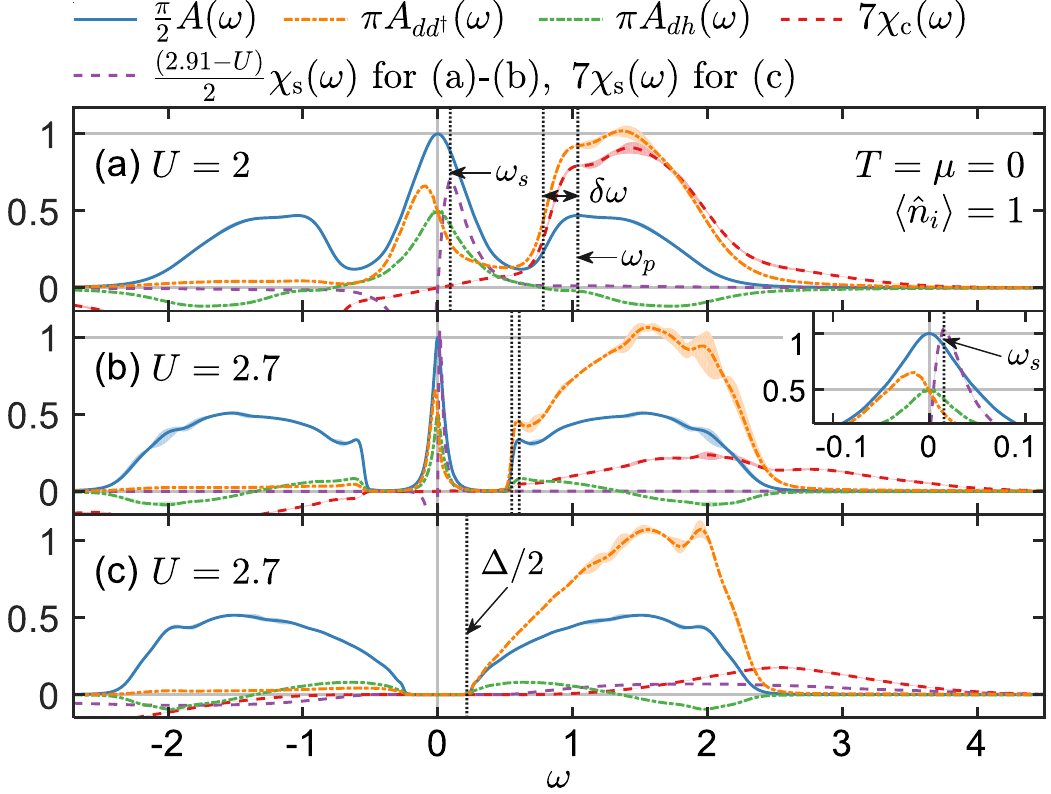}}
\caption{
Local correlation functions in (a)-(b) the metallic and (c) insulating phases of the $\mr{SU}(2)$ Hubbard model:
the local spectral function $A(\omega)$ (blue solid lines),
the correlators of doublon $d_{i\nu}$ and holon $h_{i\nu}$
operators [cf.~\Eq{eq:DHdef}] (dash-dotted lines),
charge susceptibility $\chi_c = A_{\delta \hat{n}, \delta \hat{n}}$ (red dashed lines), and
spin (i.e., flavor) susceptibility $\chi_s = A_{\vec{S},\vec{S}}/3$ (purple dashed lines), with
$\chi_{c(s)} (\omega) = -\chi_{c(s)} (-\omega)$,
$A_{dd^\dagger} (\omega) = A_{h^\dagger h}(-\omega)$, and $A_{dh} (\omega) = A_{h^\dagger d^\dagger} (\omega)$.
Here $\delta \hat{n}_i \equiv \hat{n}_i - \la \hat{n}_i \ra$
and $\vec{S}_i$ is the spin operator at site $i$.
Each correlator is averaged over different discretizations (see Sec.~I\,B of \Ref{Supp})
where the corresponding color-matched shaded area
provides an estimate for numerical uncertainties, noticeable only in the HBs.
Panels (b) and (c) show different solutions for the same value of $U$ in the coexistence regime.
In panel (b), inset zooms into the region of the QP.
We mark the location of spectral features by vertical dotted lines:
(a)-(b) subpeak position $\omega_p$ (defined as the local maximum near the inner HB edge), 
subpeak width $\delta \omega$ [defined
as the minimum positive value satisfying $A(\omega_p - \delta\omega) = A(\omega_p)/2$],
spin susceptibility peak position $\omega_s$,
and (c) inner HB edge at $\Delta/2$, where $\Delta$ is the Mott gap.
}
\label{fig:SU2}
\vspace{-.4cm}
\end{figure}

{\it DMFT+NRG.---}
We use single-site DMFT which maps the Hubbard model onto the single-impurity Anderson model (SIAM) which
provides paramagnetic solutions, by construction.
We employ the semi-circular density of states 
of the Bethe lattice with half-bandwidth $D$,
together with units $D=\hbar=k_B=1$, throughout.
We solve the SIAM by the full-density-matrix NRG 
(fdm-NRG; \cite{Weichselbaum2007,*Weichselbaum2012:mps}),
exploiting $\mr{U}(1)_\mr{charge} \otimes \mr{SU}(N)_\mr{flavor}$ symmetry~\cite{Weichselbaum2012:sym}.
\nocite{Supp,Anders2005,Anders2006,Bulla1998,Byczuk2007,Raas2009} 
The coarse-grained discretization-averaged spectral data is broadened adaptively \cite{Supp,Lee2016} 
for best possible spectral resolution at higher energies,
while preserving the intrinsic accuracy of NRG at low energies
[e.g., the Luttinger pinning~\cite{Mueller-Hartmann1989}
$\tfrac{\pi}{2} A(\omega = 0) = 1$ in the metallic phase is
accurately satisfied; see \Fig{fig:SU2}(a)-(b)].

\begin{figure}
\centerline{\includegraphics[width=.47\textwidth]{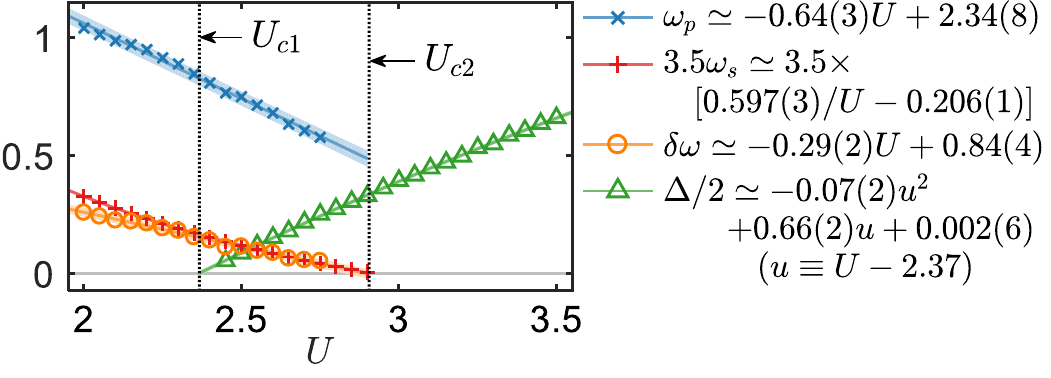}}
\caption{
The $U$-dependence of the spectral features:
the position $\omega_p$ and width $\delta \omega$ of the subpeaks, 
the peak position $\omega_s$ of spin susceptibility $\chi_s$
and the Mott gap $\Delta$ (cf.~\Fig{fig:SU2}).
Symbols are data points from the DMFT+NRG calculations,
lines are fits,
and shading gives the 95\% prediction bounds of fitting.
The zeros of the extrapolated fits of $\Delta$ and $\omega_s$ yield estimates for the critical interaction strengths $U_{c1} = 2.37(2)$ and $U_{c2} = 2.91(1)$, respectively.
}
\label{fig:DMFTfit}
\vspace{-.5cm}
\end{figure}

{\it SU(2) metallic phase.---}
We first consider the case $N=2$
equivalent to the spin-full one-band Hubbard model.
At $T = 0$ and half filling,
a metallic phase exists for $U < U_{c2} = 2.91(1)$,
and a paramagnetic insulating phase for $U > U_{c1} = 2.37(2)$.
For $U_{c1} < U < U_{c2}$ the two phases coexist
(e.g., see \Fig{fig:DMFTfit}, or Refs.~\cite{Bulla1999,Bulla2001}).

Within the metallic phase,
the local spectral function $A(\omega)$ 
features one QP and two HBs [cf.~\Fig{fig:SU2}(a)-(b)].
As $U$ increases,
the central QP narrows, the HBs widen, and the dips between the QP and the HBs deepen.
On top of this, subpeaks are present at the inner edges of the HBs,
whose position $\omega_p$ and width $\delta \omega$ decrease linearly with increasing $U$, as shown in \Fig{fig:DMFTfit}.

Local spin (i.e., flavor) and charge susceptibilities, $\chi_s$ and $\chi_c$~\cite{Raas2009:prb}, in \Fig{fig:SU2} demonstrate that the QP and the HBs of $A(\omega)$ are tied to spin and charge degrees of freedom, respectively;
that is, spin and charge excitations are energetically separated.
The peak of $\chi_s$ indicates a spin-like collective mode responsible for the QP,
which is analoguous to the Kondo resonance in the SIAM in that the spin susceptibility peaks at the Kondo energy scale~\cite{Hanl2014}.
The position $\omega_s$ and width of the $\chi_s$ peak decrease as the QP narrows with increasing $U$;
especially, $\omega_s$ has a linear dependence vs.~$1/U$, as shown in \Fig{fig:DMFTfit}.
In contrast, $\chi_c$ is suppressed within the QP region, while having long tails beyond the outer edges of the HBs.

For $T = 0^+$,
the positive and negative energy sides of a correlator
$A_{XY}(\omega)$ are derived from $\la X(t) Y(0) \ra_T$ and $\la Y(0) X(t) \ra_T = \la X^\dagger (t) Y^\dagger (0) \ra^*_T$, respectively.
Therefore the upper HB in \Fig{fig:SU2}, 
which mainly consists of $A_{dd^\dagger}$,
originates from the dynamics of the doublon
$d_{i\nu}^\dagger(0)$.
Another significant feature of $A_{dd^\dagger}$ is a peak
at $\omega = -\omega_s$.
Just after the action of $d_{i\nu} (0)$ and just before $d_{i\nu}^\dagger (t)$,
the site $i$ has only spin-$\bar{\nu}$.
Its time evolution between $0$ and $t$ with low frequency $|\omega| \simeq \omega_s$
is driven by the spin-like collective mode captured by the peak of $\chi_s$ at $\omega_s$.
In contrast,
the off-diagonal correlator $A_{dh}$ 
has a symmetric peak at $\omega = 0$. 
This reflects the particle-hole symmetric processes of destroying at the same site first a doublon and then a holon, or vice versa.
$A_{dd^\dagger}$ and $A_{dh}$
contribute comparably
to the QP, having
$A_{dd^\dagger} (0) = A_{dh} (0) = \tfrac{1}{2\pi}$.

In the metallic regime in Figs.~\ref{fig:SU2}(a)-(b)
all of the doublon and holon correlators
show peak-like features at $\omega = \pm \omega_p$.
For $U \gtrsim 2.3$ \cite{Supp}, their
contributions to these subpeaks have relative weights
$A_{dd^\dagger} (\omega_p) > A_{dh} (\pm \omega_p) > A_{dd^\dagger} (-\omega_p)$.
Our effective theory (described below) aims to reproduce 
this relative order of contributions, as well as 
the linear dependence of $\omega_p$ vs.\ $U$.

{\it SU(2) insulating phase.---}
The QP, the subpeaks, the spin-charge separation in energy space, and the peaks of the doublon and holon correlators
all disappear in the insulating phase, as depicted in \Fig{fig:SU2}(c).
Instead, a Mott gap $\Delta$ opens,
and the susceptibilities $\chi_s$ and $\chi_c$
spread over a large energy range, $|\omega | > \Delta / 2$, with suppressed heights.
While both $\omega_p$ in the metallic phase and $\Delta / 2$ in the insulating phase
correlate to the location of
the inner HB edges,
their dependences on $U$ are clearly different (see \Fig{fig:DMFTfit}).
Here the 
absence of subpeaks is consistent with previous studies~\cite{Karski2005,Karski2008,Zitko2009,Granath2012,Lu2014,Ganahl2014,Wolf2014,Ganahl2015}.
Though other works~\cite{Nishimoto2004,Gull2010,Granath2014} have reported subpeaks even in the insulating phase,
their observations are not numerically stable due to, e.g., ill-posed analytic continuation or underbroadening.

{\it DH pair interaction.---}
We will now demonstrate that the peaks of the doublon and holon correlators at $\omega = \pm\omega_p$,
which add up to the subpeaks of $A(\omega)$,
originate from a DH pair interaction within the low-energy effective Hamiltonian of the $\mr{SU}(2)$ Hubbard model.
Our theory is based on the separation of three energy scales, $\omega_s < \omega_p < U/2$, 
corresponding to the QP, the subpeaks, and the HBs, respectively.
We focus on the intermediate scale $\omega_p$ by integrating
out the larger scale $U/2$ and by approximating the physics
of the smaller scale $\omega_s$.

We first integrate out the charge fluctuation of energy
scale $U/2$, by employing a generalized SWT~\cite{Bukov2016,*Bukov2015,Lee2017a}.
We decompose the hopping term into different
components $H_v = \sum_{m = -1}^{+1}  H_{v;m}$
which cost Coulomb energy $m U$ 
since $mU H_{v;m} = [H_U, H_{v;m}]$.
Here $H_{v;0} \equiv v \sum_{\la i,j \ra,\nu}
( d_{i\nu}^\dagger d_{j\nu} - h_{i\nu}^\dagger h_{j\nu} ) +
\text{h.c.}$
describes the hopping of doublons and holons without energy cost,
whereas $H_{v;1} \equiv v \sum_{\la i,j \ra,\nu}
(d_{i\nu}^\dagger h_{j\nu}^\dagger +
 d_{j\nu}^\dagger h_{i\nu}^\dagger)$ or 
($H_{v;-1} = H_{v;1}^\dagger$)
creates (annihilates) nearest-neighbor DH
pairs by paying (gaining) energy cost $U$.
Then we write the low-energy effective Hamiltonian $H_\mr{eff}$ as  a power series in 
$v/U$,
\begin{align}
& H_\mr{eff} = H_{v;0} + H_{ss} + H_{dh} + H_\text{3-site} + O(v^3 / U^2), \label{eq:Heff} \\
& H_{ss} = \tfrac{v^2}{U} \sum_{\la i,j \ra}
4 \vec{S}_i \cdot \vec{S}_j - P_{i1} P_{j1}, \nonumber \\
& H_{dh} = \tfrac{2 v^2}{U} \sum_{\la i,j \ra} 
 (c_{j1}^\dagger c_{j2}^\dagger c_{i2} c_{i1} + P_{i2} P_{j0}) + (i \leftrightarrow j) \nonumber \\
& \phantom{H_{02}} = \tfrac{v^2}{U} \sum_{\la i,j \ra, \nu, \nu'} 
( h_{i\nu}^\dagger d_{j\nu}^\dagger + h_{j\nu}^\dagger d_{i\nu}^\dagger ) 
( d_{i\nu'} h_{j\nu'} + d_{j\nu'} h_{i\nu'} ), \nonumber 
\end{align}
where $H_\text{3-site}$ is the sum of the products of operators at three nearest neighbor sites.
The term $H_{ss} + H_{dh} + H_\text{3-site} = [H_{v;1}, H_{v;-1}]/U$, of order $O(v^2 /U)$, can be interpreted
as second-order virtual processes.
$H_\mr{eff}$ is similar to the $t$-$J$ model~\cite{Harris1967,*Chao1977,*MacDonald1988,*Eskes1994,*Eskes1994a,*Chernyshev2004},
widely used as the effective low-energy model for a Mott insulator,
but additionally contains a three-site term, $H_\text{3-site}$, and, importantly, the DH term $H_{dh}$.
Each term in \Eq{eq:Heff}
respects the $\mr{SU}(2)_\mr{charge} \otimes \mr{SU}(2)_\mr{spin}$
symmetry of the system.
See \Ref{Lee2017a} for a detailed derivation for general $N$.
Hereafter we discard the higher order $O(v^3 /U^2)$ terms.

The low-energy Hamiltonian $H_\mr{eff}$ in \Eq{eq:Heff} describes two
effective nearest-neighbor interactions
whose role and relevance depend on the phase of the system:
(i) $H_{ss}$ contains the Heisenberg spin-spin interaction.
In our paramagnetic metallic phase, this interaction induces a spin-like collective mode of energy scale $\omega_s$.
The interaction strength $v^2 / U$ is consistent with
the scaling of $\omega_s \sim 1 / U$ (cf. \Fig{fig:DMFTfit}). 
On the other hand, $H_{ss}$ becomes irrelevant
in the paramagnetic insulating phase,
where the spin susceptibility $\chi_s$ is overall suppressed.
(ii) $H_{dh}$ describes a DH pair interaction which acts on the subspace with a finite number of DH pairs. Thus $H_{dh}$ is relevant (irrelevant) in the metallic (insulating) phase.

\begin{figure}
\centerline{\includegraphics[width=.47\textwidth]{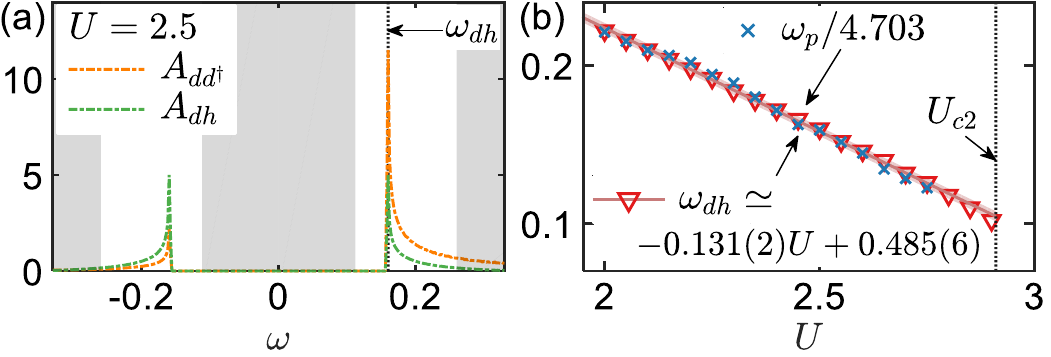}}
\caption{
(a)
Doublon and holon correlators 
$A_{dd^\dagger}$ (orange dash-dotted line) and 
$A_{dh}$ (green dash-dotted line) from our effective theory for the metallic phase.
Lower-energy spin dynamics at energies $|\omega|\lesssim \omega_s$
and higher energy scales $|\omega| \gtrsim U/2$
are neglected (as schematically indicated by the grey shading)
by employing the generalized SWT together with a
mean-field decoupling scheme.
$A_{dh}$ is symmetric, while $A_{d d^\dagger}$ is asymmetric.
Both lines have a pair of peaks at $\omega = \pm \omega_{dh}$,
showing $A_{d d^\dagger} (\omega_{dh}) > 
A_{dh}(\pm \omega_{dh}) > A_{d d^\dagger} (-\omega_{dh})$.
This is qualitatively consistent with the DMFT+NRG results for $A_{dd^\dagger}$ and $A_{dh}$ at $\omega = \pm \omega_p$
in \Fig{fig:SU2}(b)
using the same color coding.
(b) The peak position $\omega_{dh}$ from the effective theory decreases linearly with increasing $U$.
The narrow shading gives the 95\% prediction bounds of a linear fit.
$\omega_{dh}$ nicely overlaps with $\omega_p$
(data taken from \Fig{fig:DMFTfit})
up to an overall scaling factor.
We take $\Delta_{dh} = 2.91 = U_{c2}$ independent of $U$,
while the half-filled fraction $\la P_{i1} \ra$ is $U$-dependent, with the data taken from our DMFT+NRG results~\cite{Supp}.
}
\label{fig:eff}
\vspace{-.5cm}
\end{figure}

\begin{figure}
\centerline{\includegraphics[width=0.47\textwidth]{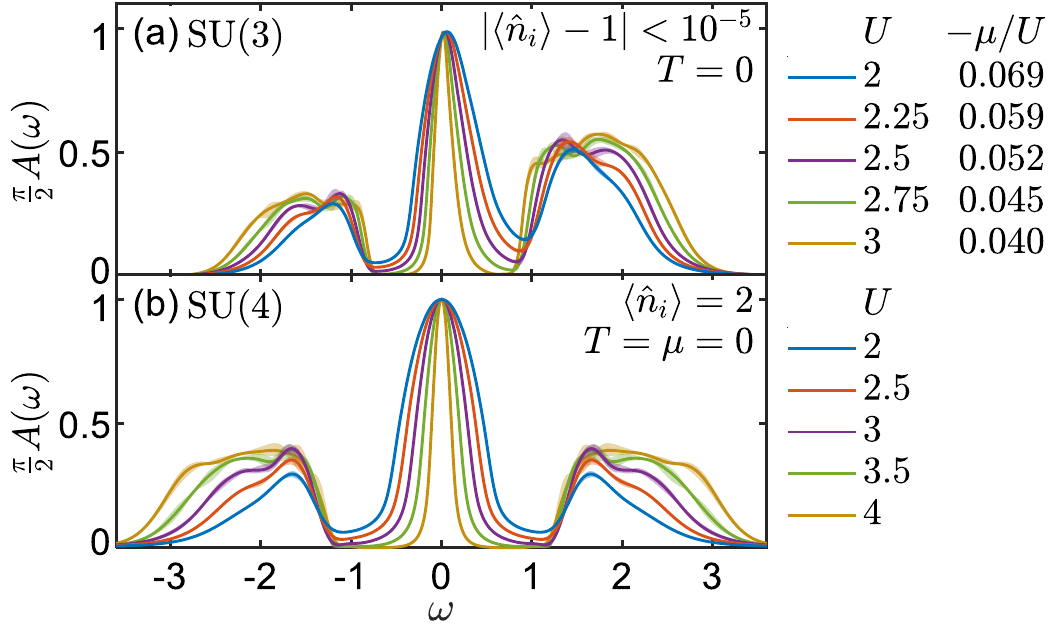}}
\caption{
Local spectral function $A(\omega)$ for (a) the $\mr{SU}(3)$ and (b) $\mr{SU}(4)$ Hubbard models in their metallic phases.
Shading again reflects the uncertainties based on discretization-averaging
(cf.~\Fig{fig:SU2}).
For $N=3$, the chemical potential $\mu$
was fine-tuned to have the integer
filling $\la \hat{n}_i \ra \simeq 1$
for different $U$, as shown in the legend of panel (a).
For $N=4$, we have $\mu = 0$ due to particle-hole symmetry.
In all cases, being in the metallic regime, subpeaks
emerge at the inner HB edges.
\vspace{-0.5cm}
}
\label{fig:SU34}
\end{figure}

{\it Doublon and holon peaks.---}
After integrating out the largest energy scale $U$,
we consider the doublon and holon dynamics governed by the effective
Hamiltonian $H_\mr{eff}$,
aiming at the intermediate energy scale $\omega_p > \omega_s$,
in the {\it metallic} phase. We simplify the physics
at lower energies ($\lesssim \omega_s$) without exactly solving $H_\mr{eff}$,
by introducing two approximations
described in detail in \Ref{Lee2017a}:
(i) We introduce a mean field, $\Delta_{dh} \equiv \tfrac{v}{2} \sum_\nu \la d_{i\nu} h_{j\nu} + d_{j\nu} h_{i\nu} \ra$,
which regards the Fermi-liquid ground state as the
``condensate'' of the DH pairs.
Then we approximate the DH interaction term as 
$H_{dh} \approx \frac{v}{U} \sum_{\la i,j \ra, \nu}
\Delta_{dh}^* (d_{i\nu} h_{j\nu} + d_{j\nu} h_{i\nu})
+ (\text{h.c.})$.
The mean-field variable $\Delta_{dh}$, comprised of
the expectation value of the pair annihilation operator
$d_{i\nu} h_{j\nu} + d_{j\nu} h_{i\nu}$,
is reminiscent of
Bardeen-Cooper-Schrieffer theory.
Here the situation is quite different, though, in that charge conservation is actually not broken, 
given that the pair annihilation operator is nothing but a summand of the decomposed hopping term $H_{v;-1}$.
The DH pairs are singlets
of the $\mr{SU}(2)_\mr{charge} \otimes
\mr{SU}(2)_\mr{spin}$ symmetry preserved in the metallic phase,
and the mean-field approximation of $H_{dh}$ also respects
that symmetry~\cite{Lee2017a}.
(ii) We decouple the doublon and holon correlators from charge and spin
density fluctations. This is
based on the numerical results that 
they are characterized by different energy scales:
charge fluctuations are suppressed in the regime $|\omega | \lesssim U/2$,
and spin fluctuations predominantly occur
at energies $|\omega| \lesssim \omega_s$ (see \Fig{fig:SU2}).
As a result, the equations of motion for the correlators close.

\Fig{fig:eff}(a) shows the resulting
doublon and holon correlators for finite $\Delta_{dh}$
in the metallic phase.
They have a pair of peaks at $\omega = \pm \omega_{dh}$, akin to their peaks at $\omega = \pm \omega_p$ in \Fig{fig:SU2}.
\Fig{fig:eff}(b) demonstrates that the DH peak position $\omega_{dh}$ from the effective theory
and the DMFT+NRG result of the subpeak position $\omega_p$ agree
well up to overall scaling factor $\simeq 4.7$
which may be
expected to arise given
the crudeness of our approximations.
In contrast, in the insulating phase $H_{dh}$ 
is irrelevant, such that $\Delta_{dh} = 0$.
As a consequence,
the subpeaks are absent in the insulating phase.

{\it Predictions for photoemission spectroscopy.---}
The QP and the HBs of the local spectral functions
have already been observed in photoemission spectroscopy~\cite{Mo2003,Sekiyama2004}.
This technique, which probes the momentum-resolved spectral function $A(\omega,\mb{k})$
(whose momentum average yields the local $A(\omega)$ discussed hitherto),
should also be able to reveal the DH subpeaks.
We have thus computed $A(\omega,\mb{k})$, see 
Figs.~S3 and S4 of \Ref{Supp}.
Our $T = 0$ results agree with prior DMFT+DMRG results from \Ref{Karski2008}, showing
that the feature in $A(\omega,\mb{k})$, which leads to the subpeak in $A(\omega)$, has distinct dispersion,
consistent with the interpretation of DH pair propagation.
Going beyond \Ref{Karski2008},
we also analyze finite $T$, and find that
the subpeak-related features survive below the critical temperature for the Mott transition~\cite{Supp}.
The distinct dispersion and $T$-dependence of the subpeak,
correlated with those of the QP,
distinguish it from other fine structure of the HBs originating
from atomic levels.
We suggest to search for such features in photoemission data,
especially in multi-band materials where the subpeaks become more pronounced, as we discuss below.

{\it $SU(N>2)$ models.---}
We also analyze the $\mr{SU}(3)$ and $\mr{SU}(4)$ Hubbard models at integer filling
$\la \hat{n}_i \ra = \lfloor N/2 \rfloor$, with the results shown in
\Fig{fig:SU34}. Similar to the case $N=2$ in \Fig{fig:SU2},
we again observe subpeaks on the inner edges of the HBs.
While the subpeaks carry small weights compared with the rest of the HBs for $N = 2$ [cf.~\Fig{fig:SU2}(b)],
the subpeaks for $N = 3, 4$ have significantly larger relative weights (cf.~purple lines in \Fig{fig:SU34}).
Even for $N = 4$, the subpeaks are clearly higher than the rest of the HBs.
Note that the QP persists more strongly at large $U \gtrsim 3$ for larger $N$, similarly to the widening of the Kondo peak in the $\mr{SU}(N)$ Kondo model~\cite{Hewson1993}.

We interpret this enhancement of the subpeaks,
as resulting from the enlarged space of 
DH pair excitations in the $\mr{SU}(N > 2)$ Hubbard models.
Generalizing the DH interaction $H_{dh}$ 
discussed above to the $\mr{SU}(N > 2)$ cases,
we find that the DH pair excitations on nearest neighbours are 3- and 15-fold degenerate in the $\mr{SU}(3)$ and $\mr{SU}(4)$ models, respectively, in contrast to the non-degeneracy in the $\mr{SU}(2)$ case~\cite{Lee2017a}.
A particularly promising area for studying this behaviour is ultracold atom physics,
where pronounced DH correlations have been reported in the 2D Hubbard model~\cite{Cheuk2016}.

{\it Conclusion.---}
We showed that the subpeaks at the inner HB edges
can be related to the effective DH pair interaction
by using a generalized SWT.
By using NRG as a real-frequency impurity solver for DMFT, 
we uncovered
detailed dynamical information on the decomposition of the local spectral function into doublon and holon correlators.
By utilizing a recently developed broadening scheme~\cite{Lee2016},
we efficiently resolved those spectral features at high energies
which had been considered challenging for the NRG
in the past due to its logarithmic coarse graining.
An effective theory based on the scale separation of
the characteristic energy scales $\omega_s$, $\omega_p$, and $U$
reproduces the linear $U$ dependence
of $\omega_p$ found numerically
in DMFT+NRG.
Our predictions should be testable using photoemission spectroscopy of correlated materials,
or in ultracold atom systems.

We thank M. Bukov, G. Kotliar, A. Mitchell, K. Penc,
A. Polkovnikov, M. Punk, and R. \v{Z}itko for fruitful discussion.
This work was supported by Nanosystems Initiative Munich.
S.L. acknowledges support from the Alexander von Humboldt Foundation and the Carl Friedrich von Siemens Foundation,
A.W. from the German Research Foundation (DFG) WE4819/2-1.

%

\clearpage

\setcounter{equation}{0}
\setcounter{figure}{0}
\setcounter{page}{1}

\renewcommand{\theequation}{S\arabic{equation}}
\renewcommand{\thefigure}{S\arabic{figure}}
\renewcommand{\thepage}{S\arabic{page}}

\centerline{\large \bf Supplementary Material}
\vspace{1em}

Here we discuss the technical details of the calculations
in the main text together with supplementary results.
In \Sec{sec:NRG}, we explain the algorithm and parameters in our NRG calculations.
In \Sec{sec:SuppRes}, we provide DMFT+NRG 
results supplementing the analysis in the main text.

The equations and figures in the Supplementary Material are referred to as numbers with S (e.g., Fig.\ S1),
while those in the main text are referred to as numbers without S (e.g., Fig.\ 1).
Also the citations occurring in the Supplementary Material refer to references given in the main text.

\section{Methods}
\label{sec:NRG}

We first summarize the procedure of the full-density-matrix NRG 
(fdm-NRG; \cite{Weichselbaum2007,*Weichselbaum2012:mps}).
Then we explain two techniques utilized in this work:
adaptive broadening~\cite{Lee2016}
for improving spectral resolution
and $\Lambda$-averaging for suppressing discretization
artifact. We also emphasize some technical aspects concerning
the application of NRG to the DMFT.

\subsection{Full-density-matrix NRG}

We consider a single-impurity Anderson model (SIAM)
in which the hybridization function between the impurity
and the bath is given by $\Gamma^\mr{in} (\omega)$.
The bath of the SIAM is discretized in energy space with
logarithmic energy grid $\pm \omega_{\max} \Lambda^{-k-z}$,
where $\Lambda > 1$ is the NRG discretization parameter,
$k\in \mbb{N}_0$ the grid index, and
$z\in ]0,1]$ the discretization shift (``$z$-shift'') with integer $n_z$.
The prefactor $\omega_{\max}$ specifies
the maximum non-zero range of the input $\Gamma^{\mr{in}} (\omega)$; see \Sec{sec:DMFT}
for its determination.
The value of $z$ is fixed for each separate NRG calculation.
The representative energy of each discretization interval
is determined by solving a differential equation~\cite{Zitko2009}.

The discretized impurity-bath Hamiltonian is tridiagonalized
to yield a semi-infinite tight-binding chain, so-called Wilson
chain. The logarithmic discretization results in an overall
exponential decay of the hopping amplitudes and the on-site
energies in the chain. Note that the on-site energies vanish
for particle-hole symmetric cases $\Gamma^\mr{in} (\omega)
= \Gamma^\mr{in} (-\omega)$.
In virtue of this exponential decay, energy scale separation
ensures that the complete set of (well approximated) energy
eigenvalues $\{ E_i^{z} \}$ and eigenstates $\{\ket{E_i^{z}}\}$
of the Wilson chain can be constructed via the iterative 
diagonalization~\cite{Anders2005,*Anders2006},
where the superscript $z$ indicates the dependence on $z$-shift.

Using the complete basis of energy eigenstates, we can compute the discrete spectrum $A_\mr{disc}^{z} (\omega)$ of general correlation function $A_{O_1, O_2} (\omega)$ in the Lehmann represesntation~\cite{Weichselbaum2007,*Weichselbaum2012:mps},
\begin{subequations}
  \begin{gather}
    A_\mr{disc}^{z} (\omega) = \sum_{ij} A_{ij}^{z} \, \delta ( \omega - \omega_{ij}^{z} ), \\
    A_{ij}^{z} \equiv \matel{E_i^{z}}{O_1}{E_j^{z}} \matel{E_j^{z}}{O_2}{E_i^{z}} (\rho_i^{z} \pm \rho_j^{z} ), \label{eq:Aijz} \\
    \omega_{ij}^{z} \equiv E_j^{z} - E_i^{z},
  \end{gather}
\end{subequations}
where $\rho_i^z = e^{-E_i^z /T} / \sum_{i'} e^{-E_{i'}^z /T}$ is the diagonal element of the density matrix at temperature $T$, $\pm$ in \Eq{eq:Aijz} takes the value $-(+)$ for (anti-)commuting operators $O_1$ and $O_2$, and we set $k_\mr{B} = \hbar = 1$.
The $\omega_{ij}^z$ values that determine the positions of the $\delta$ functions in $A_\mr{disc}^{z}$ are bunched,
and the bunches occur roughly at $\omega \sim \pm \Lambda^{-k-z}$ ($k \in \mbb{Z}$),
reflecting the logarithmic discretization of the system~\cite{Lee2016}.
To reduce discretization artefacts, we average over different choices
of $z$ and $\Lambda$, as described in more detail in \Sec{sec:Lambda} below.

The computational efficiency and accuracy in the iterative diagonalization and in computing $A_\mr{disc}^z$ can be largely enhanced, 
especially for multi-band problems, 
by exploiting non-Abelian symmetries~\cite{Weichselbaum2012:sym}.
In this work
we obtain the energy eigenstates as the multiplets of $\mr{U}(1)_\mr{charge} \otimes \mr{SU}(N)_\mr{flavor}$ symmetry.
In the iterative diagonalization, we keep up to $N_\mr{keep} = 3000$ multiplets at each step.
After the first $\lceil 2 \log_\Lambda 100 \rceil$ iterations,
we discard the multiplets with rescaled energy above $E_\mr{trunc} = 9$ for efficiency~\cite{Weichselbaum2012:mps}.

\subsection{Discretization averaging}
\label{sec:Lambda}

A drawback of the logarithmic discretization is that the {\it input}
hybridization function, $\Gamma^\mr{in} (\omega)$, is poorly resolved at
high energies; for example, each Hubbard band in \Fig{fig:SU2}
contains only two or three discretization intervals.  This is not a
problem when $\Gamma^\mr{in}$ is featureless (as in usual impurity
problems) or when the correlation functions are overbroadened at high
energies (as in previous DMFT+NRG calculations), but it can induce
some artificial features in correlation functions when
$\Gamma^\mr{in}$ has structure and the correlation functions are
finely resolved (as in our current DMFT+NRG calculations).  Since not
only physical structure but also artifacts (e.g., unphysical wriggle, overbroadening) can be self-reinforced
during the self-consistency loop, it is necessary to suppress such
discretization artifacts.

We address this problem by averaging over several discretization settings
using a combination of two different schemes.
The first, called \textit{$z$-averaging}, 
is standard practice in NRG~\cite{Zitko2009,Bulla2008}:
one averages spectral functions computed for a fixed value of $\Lambda$
over $n_z$ equally distributed $z$-shifts $z \in \{ 1/n_z, 2/n_z, \ldots, 1\}$.
The resulting discrete
energy grids all have the same spacing on a logarithmic
energy scale, but are shifted relative to each other.
This $z$-averaging is also used to improve the spectral resolution of correlation functions (see \Sec{sec:broadening} below for detail).
In addition, we use a second averaging scheme, which we
refer to as \textit{$\Lambda$-averaging}, which involves an average
over different coarse grainings on the logarithmic energy scale.

We implemented these two schemes in the following manner.
We first obtain multiple curves for the same
correlation function by independent NRG runs using sets
of different $z$ and $\Lambda$ values.
After $z$-averaging the discrete data for the
same $\Lambda$ followed by broadening, we
then average these curves over $\Lambda$.
The $\Lambda$-averaged curve of hybridization function
is fed back into the DMFT self-consistency loop
(see \Sec{sec:DMFT} below for detail).
With only two or three different
$\Lambda$'s, the discretization artifacts can be significantly
suppressed.  In this work, we average over curves computed using the
following combinations: $(\Lambda, n_z, \alpha) = (1.7, 6, 1.47)$,
$(2, 8, 1.5)$, $(2.3, 10, 1.56)$ for the $\mr{SU}(2)$ case in
\Fig{fig:SU2}, and $(\Lambda, n_z, \alpha) = (2.6, 8, 1.451)$,
$(3.2, 10, 1.49)$, $(4, 12, 1.5)$ for the $\mr{SU}(N > 2)$ case in
\Fig{fig:SU34}.  Here 
$\alpha$ is a parameter for broadening correlation functions (see \Sec{sec:broadening} below),
and the tuples $(\Lambda, n_z, \alpha)$ are chosen
to have comparable ratios of $\tfrac{\alpha}{n_z} \ln \Lambda$ for
each case.  This $\Lambda$-averaging can be used to estimate error
bars for discretization related artifacts: the shadings in
Figs.~\ref{fig:SU2}, \ref{fig:SU34}, and \ref{fig:Supp1} depict the
lower and upper bounds of the curves of different $\Lambda$'s, while
the lines show the $\Lambda$-averaged curves.

\subsection{Adaptive broadening}
\label{sec:broadening}

Physically, $A_{O_1, O_2}$ should be a continuous function of $\omega$,
since the original impurity model features a continuous bath before the discretization.
For $T = 0$, we broaden $A_\mr{disc}^{z} (\omega)$ by replacing $\delta$ functions with the log-Gaussian kernels,
\begin{equation}
\delta ( \omega - \omega' ) \Rightarrow 
\tilde{\delta}_{{\sigma}}(\omega;\omega')
\equiv \frac{\Theta(\omega \omega')}{\sqrt{\pi} {\sigma} | \omega' |}
e^{-\left( \frac{\ln | \omega / \omega'|}{{\sigma}} -
  \frac{{\sigma}}{4} \right)^2 } ,
\label{eq:logGauss}
\end{equation}
where $\sigma$ is a broadening width in log-frequency scale.
For $T \neq 0$, the broadening kernels are modified from \Eq{eq:logGauss}, see \Ref{Lee2016} for details.
Based on the observation on the bunching of $\delta$ functions roughly at $\omega \sim \Lambda^{-k-z}$,
the conventional broadening scheme uses constant $\sigma$ for all spectral contributions,
i.e., the broadening width for a weight $A_{ij}^z$ in linear-frequency scale is simply proportional to $|\omega_{ij}^z|$.
Thus the spectral features at high energies (e.g., side peaks) are generally overbroadened by the conventional scheme;
the bunches contributing to such features are distributed in an ``irregular'' manner that does not follow the $\Lambda^{-k-z}$ pattern.

A recipe to improve the spectral resolution within the conventional broadening scheme is $z$-averaging (see also above),
where one broadens the discrete data averaged over different NRG calculations of $A_\mr{disc}^{z}$ for different $z$-shifts,
$\bar{A} (\omega) = \tfrac{1}{n_z} \sum_z A_\mr{disc}^{z} (\omega)$.
Since the $\delta$-function bunches at $\omega \sim \Lambda^{-k-z}$ for a given $z$-shift interlace with the bunches for the other $z$-shifts,
one can use a narrower width, $\sigma \propto 1/n_z$, to achieve the continuity.
Then the resolution improves as $n_z$ increases, but the improvement is limited~\cite{Lee2016}.

Recently, two of the authors developed an adaptive broadening scheme~\cite{Lee2016} which further enhances the spectral resolution at high energies, in combination with $z$-averaging.
The adaptive scheme broadens each discrete peak in $\bar{A} (\omega)$ with individual width:
$\delta (\omega - \omega_{ij}^z)$ is broadened to $\tilde{\delta}_{\sigma_{ij}} (\omega; \omega_{ij}^z)$, where
\begin{equation}
{\sigma}_{ij} = \frac{\alpha}{n_z} \frac{d \ln | \omega_{ij}^z |}{d z}
= \frac{\alpha}{n_z |\omega_{ij}^z|} \left| \frac{d E_j^z}{d z} - \frac{d E_i^z}{d z} \right|,
\label{eq:sigmaij}
\end{equation}
and $\alpha$ is overall prefactor of order $O(1)$.
Here $\sigma_{ij}/\alpha$ estimates the distance on log-frequency scale,
from one bunch of $\delta$ functions at $\omega_{ij}^z$ for a given $z$-shift to its neighbouring bunch at $\omega_{ij}^{z + (1/n_z)}$ for the next $z$-shift;
this estimate captures the irregular distribution of $\delta$-function bunches contributing to the spectral features at high energies,
and assigns smaller broadening widths $\sigma_{ij}$ for these bunches.

As a result, the adaptive scheme better resolves such features, and the enhancement is more significant for larger $\Lambda$ or smaller $n_z$.
Of course, the adaptive scheme retains the intrinsic accuracy of NRG at low energies;
the Friedel sum rule and the Luttinger pinning are fulfilled with sub-1\% error.
Note that, in this work, we use the lower bound $\sigma_{ij} \geq (\ln \Lambda)/15$ to avoid unnecessary underbroadening.

\subsection{DMFT}
\label{sec:DMFT}

We use single-site DMFT based on the semi-circular
density of states 
$\rho_0 (\omega) = \tfrac{2}{\pi D^2}\sqrt{D^2 - \omega^2}$
for the non-interacting lattice, 
corresponding to a
Bethe lattice of coordination number $z \to \infty$ and
$v \propto 1/\sqrt{z}$.
We set the half-bandwidth $D \equiv 2 v \sqrt{z} := 1$
as the unit of energy, as well as $\hbar = 1$ throughout.

In the coexistence region of the metallic and insulating phases (for $U_{c1} < U < U_{c2}$ in our Hubbard models at $T = 0$),
the metallic (insulating) choice of $\Gamma^\mr{in}$ as the initial seed in the self-consistency loop results in a metallic (insulating) solution.
As the metallic and insulating initial seeds, we chose the metallic solution
$\pi D^2 \rho_0 (\omega) / 4 = \sqrt{D^2 - \omega^2} / 2$ (which is the
exact solution for $U = 0$),
and the insulating solution for $U = 3.2D$ (obtained by using the $U=0$
solution as the initial seed), respectively.

In each iteration (except for the first iteration), 
$\Gamma^\mr{in} (\omega)$ is determined by the result from the previous iteration that
has exponentially decaying tails at large frequencies;
we define $\omega_\mr{max}$ as the largest energy satisfying $\Gamma^\mr{in} (\omega_\mr{max}) = \mr{max} (\Gamma^\mr{in}) / 100$, without loss of generality.
Then we define the discretization grid $\pm \omega_{\max} \Lambda^{-k-z}$ as mentioned above.
We compute the impurity self-energy as the ratio of two correlation functions, $\Sigma_\mr{imp} \equiv \la [ c_{\mr{imp},\nu} , \tfrac{U}{2} n_\mr{imp} (n_\mr{imp} - 1) ] || c_{\mr{imp},\nu}^\dagger \ra_\omega / \la c_{\mr{imp},\nu} || c_{\mr{imp},\nu}^\dagger \ra_\omega$~\cite{Bulla1998}, where 
$\la X || Y \ra_\omega = G_{X Y} (\omega)$ is the 
retarded correlation function,
$c_{\mr{imp},\nu}$ annihilates a particle of flavor $\nu$ at the impurity, and $n_\mr{imp} = \sum_\nu c_{\mr{imp},\nu}^\dagger c_{\mr{imp},\nu}$.

As we consider the semi-elliptic density of states $\rho_0 (\omega)$ of the lattice,
the local spectral function $A(\omega)$ and the hybridization function $\Gamma^\mr{out} (\omega) = \pi D^2 A(\omega) / 4$ for the next iteration
can be derived from $\Sigma_\mr{imp}$ via an explicit relation
$A(\omega) = \tfrac{-2}{\pi D^2} \im ( \xi - \sqrt{ \xi^2 - D^2 } )$
without numerical integration, where $\xi \equiv \omega + \mu - \Sigma_\mr{imp}$.
We continue the loop until the self-consistency criterion $|\Gamma^\mr{in} - \Gamma^\mr{out}| < \pi / 10^{3}$ is satisfied.

The above relation between $\Sigma_\mr{imp}$ and $A(\omega)$, which originates from the semi-ellipticity of $\rho_0$, 
leads to two interesting properties at self-consistency, when $\Gamma^\mr{in} = \Gamma^\mr{out}$:
(i) the impurity Green's function $( \xi - \Gamma^\mr{in} )^{-1}$, improved by using self-energy~\cite{Bulla1998},
is equivalent to the local lattice Green's function $\tfrac{2}{D^2} (\xi - \sqrt{\xi^2 - D^2})^{-1}$.
(ii) Since $\Sigma_\mr{imp} = \omega + \mu - (\la c_{\mr{imp},\nu} || c_{\mr{imp},\nu}^\dagger \ra_\omega)^{-1} - \tfrac{D^2}{4} \la c_{\mr{imp},\nu} || c_{\mr{imp},\nu}^\dagger \ra_\omega$,
the kinks (i.e., big changes in the first derivatives) of $\Sigma_\mr{imp}$~\cite{Byczuk2007,*Raas2009} and $A_{cc^\dagger} \equiv \tfrac{-1}{\pi} \im \la c_{i\nu} || c_{i\nu}^\dagger \ra_\omega$ are directly related.
Therefore, the sharp peaks of $A_{d d^\dagger} (-\omega_s)$ and $A_{h^\dagger h} (\omega_s)$ [cf.~\Fig{fig:SU2}]
result in the kinks in $A_{cc^\dagger}$ as well as those in $\Sigma_\mr{imp}$.

\begin{figure}
  \centerline{\includegraphics[width=.49\textwidth]{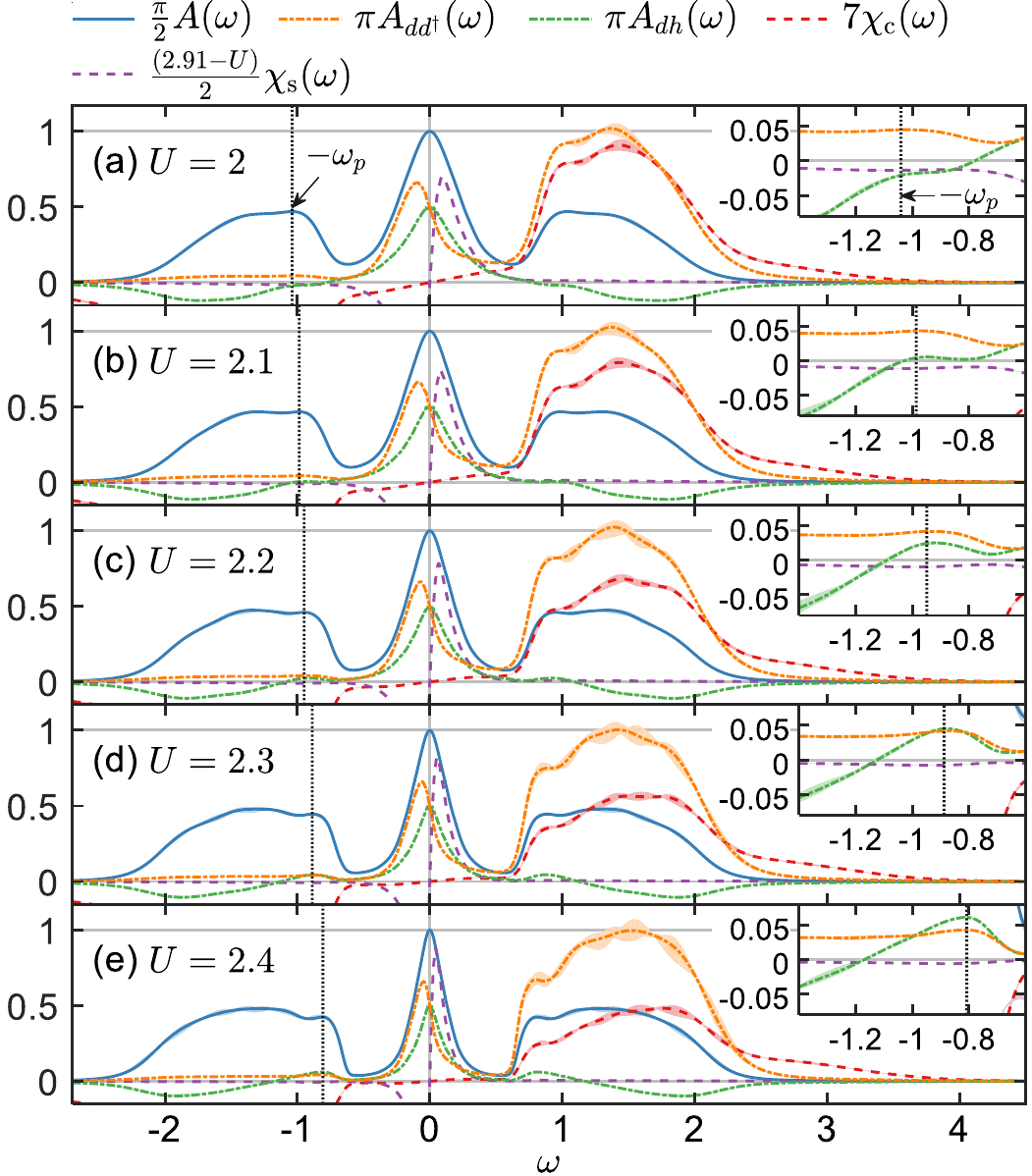}}
  \caption{ 
    Local correlation functions for $2 \leq U \leq 2.4$ in the $\mr{SU}(2)$ model.
    Refer to the caption of \Fig{fig:SU2} in the main text for the description of lines and shades.
    Insets zoom into the region $\omega \sim -\omega_p, -U/2$, and the vertical dotted line indicates the location of the left subpeak $-\omega_p$.
  }
  \label{fig:Supp1}
\end{figure}

\begin{figure}
  \centerline{\includegraphics[width=.49\textwidth]{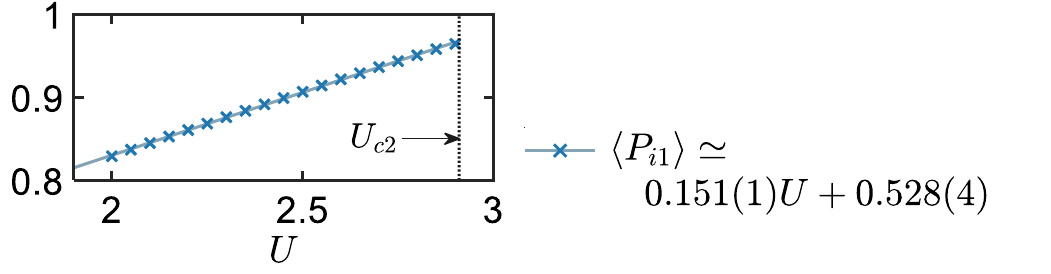}}
  \caption{
    Probability of single occupation $\la P_{i1} \ra$ vs $U$ of the $\mr{SU}(2)$ Hubbard model.
    Symbols are data points from the DMFT+NRG calculations and line is fit.
    The 95\% prediction bounds of fitting is invisibly narrow.
  }
  \label{fig:Supp2}
\end{figure}

\section{Supplementary results}
\label{sec:SuppRes}

\Fig{fig:Supp1} shows
how the correlation functions in the metallic phase
change for $2 \leq U \leq 2.4$.
This is similar to \Fig{fig:SU2}, except for the narrower
range in $U$ and the zooms into $\omega \sim -\omega_p$ in the insets.
The peak of the off-diagonal correlators,
$A_{dh} (\pm \omega_p)$ or $A_{h^\dagger d^\dagger} (\pm \omega_p)$,
grows with increasing $U$;
while it is only a shoulder for $U \lesssim 2$, for $U \gtrsim 2.3$
it becomes a peak that actually exceeds
the shallow peak of the diagonal correlators,
$A_{dd^\dagger} (-\omega_p)$ or $A_{h^\dagger h} (\omega_p)$.
Note that the off-diagonal correlators are negative
for $|\omega| \gtrsim U/2$ for $U \gtrsim 2.1$
and their full integrals are zero by sum rule.
In contrast, the diagonal correlators are 
positive, throughout, by construction.

\Fig{fig:Supp2} shows the data for the probability
of single occupation $\la P_{i1} \ra$ which entered
the mean-field decoupling analysis in \Fig{fig:eff}
in the main text.
It shows a clear linear dependence on $U$.

\begin{figure}
\centerline{\includegraphics[width=.49\textwidth]{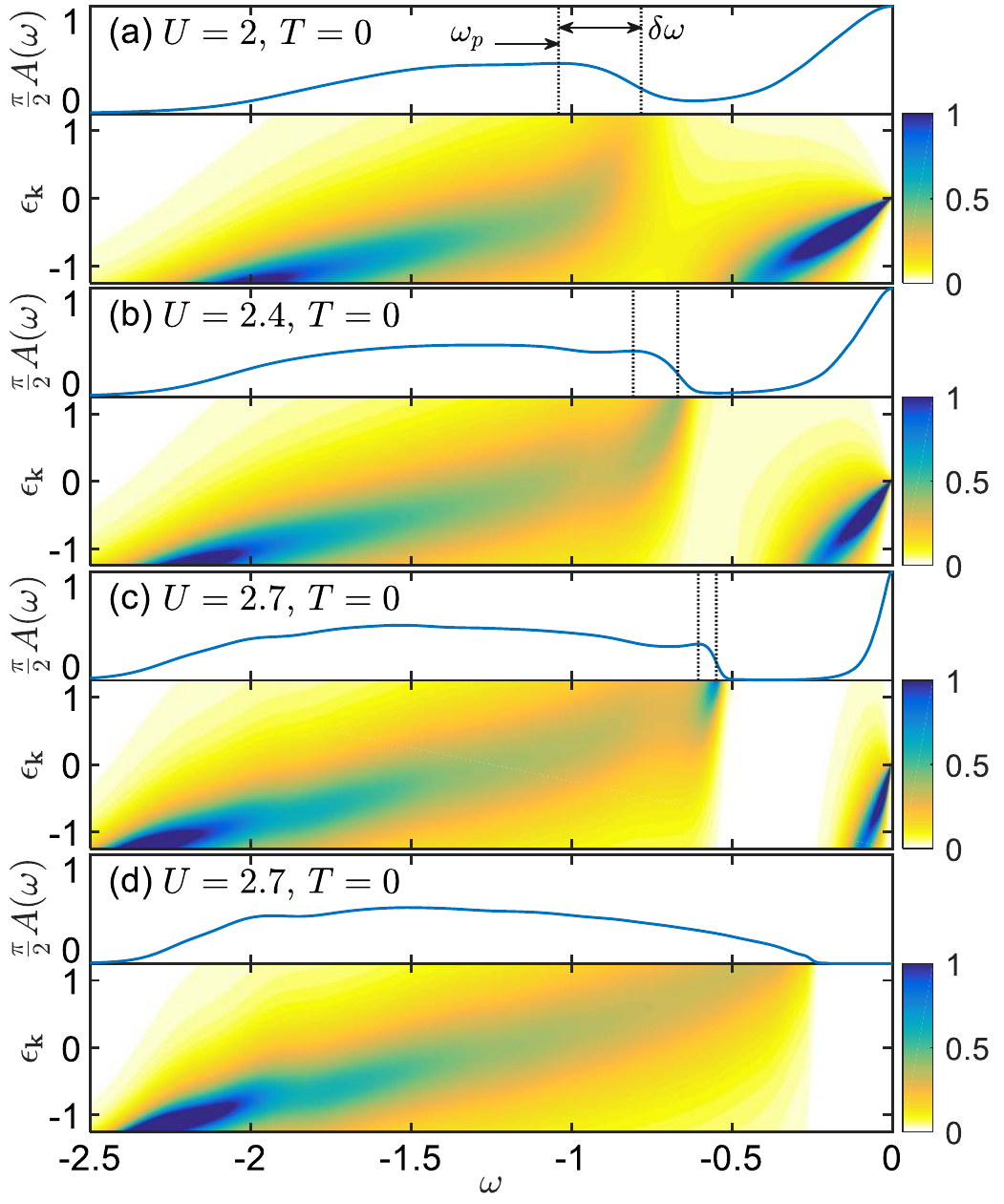}}
\caption{
Local spectral function $A(\omega)$ (upper line plots) and 
momentum-resolved spectral function $A(\omega, \epsilon_\mb{k})$ (lower color plots) in 
(a)-(c) the metallic and (d) insulating phases of the half-filled $\mr{SU}(2)$ Hubbard model,
for different values of $U$ at $T = 0$.
For panels (a)-(c) we initiated the DMFT loop with metallic local spectral function,
while for panel (d) with an insulating one; see \Sec{sec:DMFT} for detail.
Hence panels (c) and (d) 
lie in the coexistence regime, which exhibits two
different phases for the same value of $U = 2.7$.
The range, i.e., position and width, of the subpeak
in panels (a)-(c) are indicated by vertical dotted lines.
}
\label{fig:Akw_U}
\end{figure}

\begin{figure}
\centerline{\includegraphics[width=.49\textwidth]{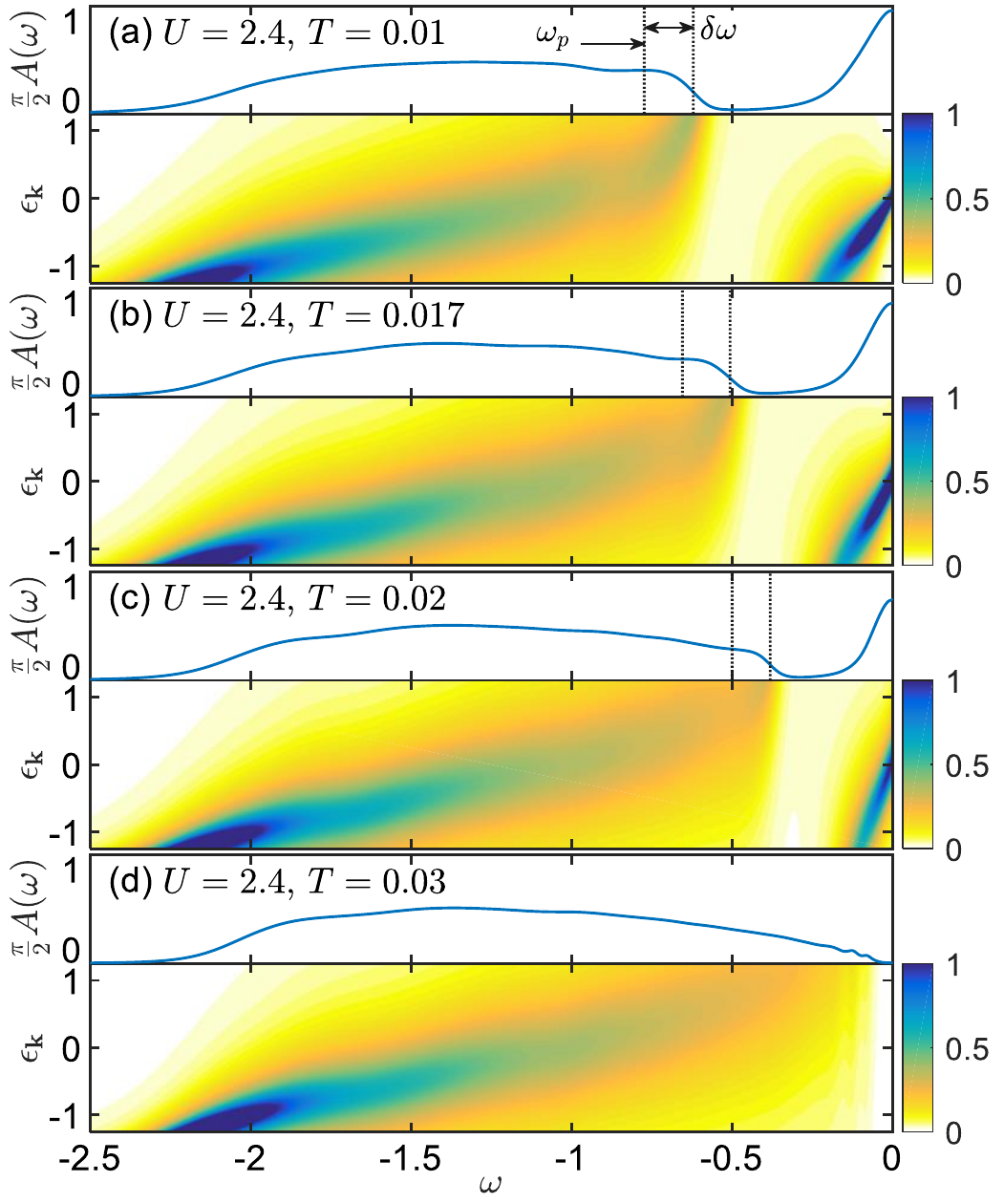}}
\caption{
Similar analysis
as in \Fig{fig:Akw_U}, but for constant $U = 2.4$
while changing $T$.
For all panels, we initiated the DMFT loop with metallic local spectral function; see \Sec{sec:DMFT} for detail.
}
\label{fig:Akw_T}
\end{figure}

Up to now, we have shown the local (i.e., momentum-averaged) correlation functions at $T = 0$.
In photoemission spectroscopy experiments,
the spectral functions are usually measured resolved in momentum $\mb{k}$ at finite $T$.
Therefore we have also studied
the momentum-resolved spectral functions,
\begin{equation}
A(\omega, \epsilon_\mb{k}) = \frac{-1}{\pi} \, \im \frac{1}{\omega + \mu - \epsilon_\mb{k} - \Sigma (\omega) + i0^+} ,
\nonumber
\end{equation}
with the results presented in Figs.~\ref{fig:Akw_U} and \ref{fig:Akw_T}
for the half-filled $\mr{SU}(2)$ Hubbard model
for different values of $U$ and $T$.
In the DMFT, since the self-energy $\Sigma$ is approximated to be independent of $\mb{k}$,
the $\mb{k}$-dependence of the spectral function appears only as the dependence on the non-interacting single-particle energy $\epsilon_\mb{k}$ with momentum $\mb{k}$, i.e., $A(\omega,\mb{k}) = A(\omega,\epsilon_\mb{k})$.
For comparison, we also plot the local spectral function $A(\omega)$,
which is the average of $A(\omega,\epsilon_\mb{k})$ over momentum space,
\begin{equation}
A (\omega) = \int \mr{d}\epsilon \, \rho_0 (\epsilon) A(\omega,\epsilon) ,
\nonumber
\end{equation}
where $\rho_0 (\epsilon) = \tfrac{2}{\pi D^2} \sqrt{D^2 - \epsilon^2}$
is the density of states for non-interacting lattice
considered for our DMFT calculations.
Here we plot only the spectral functions for negative frequencies,
since the photoemission spectroscopy mainly accesses the energy below the Fermi level.
At half filling,
one has for positive frequencies by particle-hole symmetry:
$A(\omega) = A(-\omega)$ and
$A(\omega, \epsilon_\mb{k}) = A(-\omega,-\epsilon_\mb{k})$.

\Fig{fig:Akw_U} illustrates how $A(\omega, \epsilon_\mb{k})$ evolves with increasing $U$.
For the metallic results in panels (a)-(c),
there are two most pronounced ridges of spectral intensity:
First, one ridge extends from the origin at $(\omega,\epsilon_\mb{k}) = (0, 0)$.
By parametrizing its ridge line as $\omega_0 (\epsilon_\mb{k})$,
we find an approximately linear dispersion,
$\omega_0 \simeq Z \epsilon_\mb{k}$.
The slope $Z = m_0/m^*$ corresponds to the inverse
effective mass of the quasiparticle [equivalently,
$Z$ is the quasiparticle weight proportional to the peak position $\omega_s$ of the local spin susceptibility,
i.e., $Z \approx 3 \omega_s$].
As $U$ increases, the effective mass $m^*$
diverges at $U = U_{c2}$.
Therefore the slope $\mr{d}\omega_0 / \mr{d}\epsilon_\mb{k}$
goes to zero, as the QP disappears
entirely.
Second, a broad ridge is associated with
the lower HB in $A (\omega)$
which stretches over a wide
range of energies $\omega \lesssim -D$.

In Fig.~\ref{fig:Akw_U}(b)-(c),
another intermediate ridge, whose ridge line is parametrized as $\omega_1(\epsilon_\mb{k})$,
appears in the region $\omega_1 \simeq -\omega_p$ and $\epsilon_\mb{k} \lesssim 1$,
which contributes to a subpeak in $A(\omega)$
at $\omega = -\omega_p$.
As $U$ increases within the metallic phase, 
this ridge becomes more pronounced, sharper, and better separated
from the other ridges.
On the other hand, in the insulating phase as in panel (d),
both, the ridges for the subpeak and quasiparticle,
disappear. Therefore the occurance or not
of this intermediate ridge is completely tied to
the behavior already seen in the spectral function
$A(\omega$) itself.

Note that this result is consistent with the DMFT+DMRG calculation for $T = 0$
by Karski {\it et al.}~\cite{Karski2008}.
Compared to the DMFT+DMRG results of \Ref{Karski2008},
the spectral resolution achieved here by DMFT+NRG is better at low energies and similar at high energies.
While the former is to be expected, the latter is not,
due to the reliance of NRG on logarithmic discretization.
Here we nevertheless achieve a rather high resolution even at high energies by using the refined broadening scheme developed in \Ref{Lee2016}. 

The range of the dispersion of the intermediate ridge
is comparable with the subpeak width $\delta\omega$ in $A(\omega)$.
It is also consistent with the result in \Fig{fig:DMFTfit}
where $\delta\omega \approx 3.5 \omega_s$ for large $U > 2.5$.
The slope $\mr{d}\omega_1/\mr{d}\epsilon_\mb{k}$ 
for larger $\epsilon_\mb{k}$ on the order of the
half-bandwidth $D=1$
is roughly half of the quasiparticle weight
$Z \simeq \mr{d}\omega_0 / \mr{d}\epsilon_\mb{k}$,
which suggests that the underlying object responsible
for the subpeaks has
about twice the effective mass of an individual quasiparticle.
In this sense our NRG-based numerical results are
consistent with our interpretation of the subpeaks
as arising from doublon-hole pairs.

The temperature dependence of the momentum-resolved
spectral function $A(\omega, \epsilon_\mb{k})$ is
analyzed in \Fig{fig:Akw_T}.
As $T$ increases,
the subpeak in $A(\omega)$ and the subpeak ridge in $A(\omega,\epsilon_\mb{k})$ become
suppressed and blurred,
in accordance with the suppression of the
quasiparticle-related spectral features.
This is consistent with our effective theory that
the subpeaks originate from the doublon and holon
excitations on top of the Fermi-liquid ground state,
which serves as as the DH condensate.
When the Fermi-liquid quasiparticles become ill-defined,
also the DH condensate breaks down.
Despite such thermal suppression,
we emphasize that the subpeak-related features are still visible 
at temperatures as large as $T \lesssim 0.02$,
i.e., close to the critical temperature is $T_c (U = 2.4) \simeq 0.025$.

\end{document}